\documentclass[%
 reprint,amsmath,amssymb, 
prstab,
]{revtex4-2}
\usepackage{graphicx}
\usepackage{dcolumn}
\usepackage{bm}
\usepackage[dvipsnames]{xcolor}
\newcommand{\be}{\begin{equation}}
\newcommand{\ee}{\end{equation}}
\usepackage{ulem}
\usepackage{siunitx}

\newcommand{\ie}{\textit{i.e.}}
\newcommand{\eg}{\textit{e.g.}}
\newcommand{\rms}{\textit{rms}}
\newcommand{\etal}{\textit{et al}.}

\newcommand{\Smilei}{{\sc Smilei}}
\begin{document}

\title{Accurate electron beam phase-space theory for ionisation injection schemes driven by laser pulses}
\author{Paolo Tomassini$^{1}$,  Francesco Massimo$^{2}$, Luca Labate $^{1,3}$ and Leonida A. Gizzi$^{1,3}$ }

\affiliation{$^{1}$Intense Laser Irradiation Laboratory, INO-CNR, Via Moruzzi 1 Pisa (Italy)}
\email{paolo.tomassini@ino.it}
\affiliation{$^{2}$Maison de la Simulation, CEA, USR 3441 Bâtiment 565 – Digiteo, 91191 Gif-sur-Yvette cedex  (France)} 
\affiliation{$^{3}$INFN, Sect. of Pisa,   Largo Bruno Pontecorvo 3 (Italy)}

\date{\today}

\begin{abstract}
 After the introduction of the ionization-injection scheme in Laser Wake Field Acceleration and of related high-quality electron beam generation  methods as two-color or  the Resonant Multi Pulse Ionization injection, the theory of thermal emittance by C. Schroeder \etal\, has been used to predict the beam normalised emittance obtainable with those schemes. In this manuscript we recast and extend such a theory, including both higher order terms in the polinomial laser field expansion and non polinomial corrections due to the onset of saturation effects in a single cycle. Also, a very accurate model for predicting the cycle-averaged $3D$ momentum distribution of the extracted electrons, including saturation and multi-process events, is proposed and tested. We show that our theory is very accurate for the selected processes of Kr$^{8^+\rightarrow10^+}$  and Ar$^{8^+\rightarrow10^+}$, resulting in a a maximum error below $1\%$ even in deep saturation regime.  This highly accurate prediction of the beam phase-space can be implemented \eg\, in laser-envelope Particle in Cell (PIC) or hybrid PIC-fluid codes, to correctly mimic the cycle-averaged momentum distribution without the need of resolving the intra-cycle dynamics. Finally, we introduce further spatial averaging with Gaussian longitudinal and transverse laser profiles, obtaining expressions for the whole-beam emittance that fits with Monte Carlo simulations in a saturated regime, too.
\end{abstract}

\maketitle


\section{Introduction}
In the past decades, many injection schemes of electron beams in the accelerating wakefield excited by laser pulses \cite{TajimaDawson79,Malka2002,Esarey2009,Malka2012} have been proposed and tested. Among them, injection by background density variation \cite{bulanov1998particle,suk2000plasma,hosokai2003effect,tomassini2003production,Schmid2010,Buck2013,Wang2016,Swanson2017}, collinear colliding pulses injection \cite{Faure2006,Rechatin2009,Hansson2016colliding} and multi-pulse ionization injection schemes as Two Color ionization injection \cite{Yu2014,Schroeder2018} and Resonant Multi-Pulse Ionization injection (ReMPI) \cite{tomassini2017,tomassini2019high1,tomassini2019high2} are very promising in terms of transverse beam quality, being capable of generating electron beams with normalized emittances as low as tens-of-nm,  as shown by analytical results and numerical simulations.

Accuracy of numerical simulations of ionization injection processes can be extremely challenging when schemes providing good-quality beams are investigated, as those required to accelerate electron bunches suitable to drive X-ray Free Electron Laser for the EuPRAXIA project \cite{assmann2020eupraxia}  or similar projects based on high gradient plasma accelerator \cite{albert20212020}. This is because the longitudinal grid spacing should be small enough to efficiently resolve the extraction process, occurring in a tiny fraction (usually $\approx 1/5$) of the ionization pulse wavelength. The use of reduced envelope models in conjunction with analytical models to correctly mimic the newborn electrons phase-space (\eg \thinspace\thinspace QFluid \cite{Tomassini2015MatchingSF,tomassini2017}, INF$\&$RNO \cite{Benedetti2010}, ALaDyn \cite{ALaDyn2008,Terzani2019} and \Smilei\thinspace   \cite{Smilei2018,massimo2020})  can therefore be advantageous when long and large grid-size simulations are needed. In this respect, highly accurate analytical predictions of the \rms\, transverse momentum, or even more accurate models for the phase-space distribution of the extracted electrons  are needed. In a seminal paper of 2014, C. Schroeder \etal\thinspace \cite{schroeder2014thermal} set for the first time a comprehensive theory of ionization injection thermal emittance with a  single laser pulse. This theory is currently used in the codes cited above and constitutes the state-of-the-art of the analytical results for single pulse ionization injection schemes, to the authors' knowledge.

In the following, we will suppose that the ionization laser pulse of amplitude $a_{0}$, with  polarization axis $x$ and carrier wavelength $\lambda_0$ is propagating along positive $z$. Its amplitude is large enough to provide an electric field above the ionization threshold for the tunnel field-ionization process. 
Once electrons are extracted from the ions, their dynamics follow the prescription for a generic charged particle in an (almost) plane-wave laser pulse. After averaging the momenta during the whole first laser pulse oscillation, we obtain the initial cycle-average normalised $3D$ momentum $\vec{u}=\vec{p}/m_ec$  (see  \cite{massimo2020} and references therein)
\be
\label{eq:initialmomenta}
\bar{u}_x=-a_{0,e}\sin(\xi_e)\, \, , \bar{u}_y=0\,\, ,
\bar{u}_z=\frac{1}{2}a_{0,e}^2\left[\sin^2(\xi_e)+\frac{1}{2}\right]\,, 
\ee
where   $\xi_e$  is  the ionization  pulse phase at the extraction time and $a_{0,e}$ is the local normalised pulse amplitude at the extraction position. As the electrons slip-back in the laser field, their quivering decreases, while the longitudinal ponderomotive force gradually also reduces the cycle-averaged longitudinal momentum $\bar{u}_z$. Finally, as the pulse completely overpasses the particle, the 3D {\it residual} momentum 
\be
\label{eq:momenta}
u_x=\bar{u}_x\, \, , u_y=\bar{u}_y\,\, ,
u_z=\bar{u}_z-\frac{1}{4}a_{0,e}^2\,
\ee 
can be evaluated by neglecting transverse ponderomotive effects and pulse evolution during the slippage.  It is worth to note here that, while the (initial) cycle-averaged momentum in Eq. \ref{eq:initialmomenta} is used in \eg\, {\it envelope} ionisation models, the residual momenta of Eq. \ref{eq:momenta}  can be employed, in conjunction with the transverse residual position estimate, to evaluate the minimum normalised emittance of the extracted bunch, as in \cite{schroeder2014thermal}. In this paper, the effect of the plasma wakefield either for the particle trapping (see \cite{zhidkov2020characterization} for a detailed study) or beam emittance growth due to the (possible) presence of non linear transverse forces are not taken into account. 

Theory from C. Schroeder \etal\thinspace \cite{schroeder2014thermal} also  shows that, in the optimal conditions of unsaturated ionization, the newborn electrons are extracted in tiny slabs centered at the maxima of the electric field strength $E=|\vec{E}(\vec{x},t)|$.  For a given position, and after having defined the phase of $E=E_0|\cos(\xi)|$ so as $\xi=0$ corresponds to a given maximum of $E$, the analytical theory shows that the local particle extraction phase $\xi_e$ shows a Gaussian distribution around $\xi_e=n\pi$, with $n$ integer, and variance $\sigma_\xi\simeq \Delta$ (note that in Ref. \cite{schroeder2014thermal} the phase extraction variance is named $\sigma_\psi$), where
\begin{equation}
\label{eq:Delta}
\Delta=\left(\frac{3E_0}{2E_a}\right)^{1/2}\cdot \left(\frac{U_H}{U_I}\right)^{3/4}.
\end{equation}
Here  $E_0$ is the ionization pulse strength, $E_a\simeq 0.51TV/m$ is the atomic field strength, $U_{H,I}$ are the ionization potentials of hydrogen and of the atomic selected level to be ionized, respectively. Consequently, the \rms\, residual  particle momentum $\sigma_{u_x}=\sqrt{\langle( u_x)^2\rangle}$ along the ionization pulse polarization is approximately  $a_{0}\Delta$.  
High-quality electron bunches are obtained by minimizing the transverse \rms\, momentum and this is accomplished by a minimization of $\sigma_\xi$, which should assume the lowest possible value compatible with the possibility of extracting the electrons from the selected atomic level of the dopant atoms. As an example, N$^{5^+\rightarrow6^+}$, Ar$^{8^+\rightarrow9^+}$ and $Kr^{8^+\rightarrow9^+}$ transitions are usually employed in ReMPI or Two-Color schemes. The optimal values of $\Delta\simeq\sigma_\xi$ for those processes are of about 0.29, 0.24 and 0.22, respectively (see below). 

The possibility of using very accurate predictors of the  \rms\,  normalised emittance along the polarisation axis for either particles extracted in a single cycle or by the whole laser pulse is of paramount importance for High-Quality beam production studies. Moreover, as standard requests refer to {\it both} high-charge and high-quality for the beam, working points in a saturated or partially saturated regime are often selected. Motivated by the needs reported above, we recast the theory in \cite{schroeder2014thermal} for the local and global bunch parameters, so as to include all the relevant terms of order $\Delta^2$, and to include additional $\Delta^4$ terms. In this work, we addressed the need of high accuracy \rms, predictors in the unsaturated regime, with errors between analytical results and numerical simulations below $1\%$ (see Sects. IIIa and IVa). As high-charge beams are needed, however, higher pulse amplitudes are used so as to extract more charge, therefore exploring partially or even fully saturate regimes. There, a gradual increase of the global normalised emittance is found by simulations, as already pointed out in \cite{schroeder2014thermal}. Our analytical theory that includes global saturation effects confirms the emittance increase and very accurately fits the simulation results (see Sec. IVb).  Moving with increasingly higher amplitudes, we explore the saturation limit {\it within a single laser cycle}.    The phase space of the electrons extracted in a single-cycle saturated  regime (see Sect. IIIb) reveals fine structures that may help the understanding of either experiments \cite{guenot2017relativistic,Faure2019} or PIC simulations \cite{lifschitz2012optical}  results when high intensity, very short pulses are used. Our model for the phase-space reveals to be extremely accurate in this regime, too (see Sect. IIIb) and predicts a {\it reduction} of a transverse momentum once the fully  saturated regime is reached. Very large pulse amplitudes, however, may lead to switch-on {\it multiple} ionisation stages. In this work we  also propose an accurate model for this double-ionisation process (see Sect. IIIc).

\section{Setting up the pulse amplitude for tunnel ionization}
In the following, the tunnel ionization process occurring in a (single) laser field is considered.  The instantaneous ionization rate can  be  described by the ADK formula \cite{Perelomov1966,ADK1986,Yudin2001,Nuter2011}
expressed in terms of the electric field normalised to the critical ADK field $\rho_0\equiv \frac{3E}{2E_a}\left(\frac{U_H}{U_I}\right)^{3/2}=a_0/a_c$ (here $a_c\simeq 0.107 \lambda_0\left(\frac{U_I}{U_H}\right)^{3/2}$ ), introduced in \cite{tomassini2017}: 
\begin{eqnarray}
\label{eq:ADK}
\frac{d n_e}{dt}&=&W \cdot (n_{0,i}-n_e)\, ,\nonumber \\
W&=& C \left(\rho_0|\cos\xi|\right)^\mu\exp\left(-\frac{1}{\rho_0|\cos\xi|}\right)
\end{eqnarray}
where $n_e$ is the number of extracted particles and $n_{0,i}$ is the initial number of available ions,  $C$ depends of the atom species and ionization level (there are some different versions for $C$ \eg\, Eq. 6 in \cite{tomassini2017}). The exponent $\mu$ in (\ref{eq:ADK}) is defined as 
\be 
\mu=-2n^*+|m|+1\, ,
\ee
being  $n^*=Z\sqrt{U_H/U_I}$ and $m$ the effective principal quantum number of the ion with final charge $Ze$ and the projection of the angular momentum, respectively. The peak normalised amplitude $\rho_0=a_0/a_c$ is related to the $\Delta$ term in \cite{schroeder2014thermal} as $\rho_0=\Delta^2$. The evaluation of the number of extracted electrons and spatial averages of $\sigma_{u_x}$ will be strongly simplified by expressing the average ionization rate over a single ionization pulse cycle $\langle W(\rho_0)\rangle$ as:
\begin{eqnarray}
\label{Wmean}
\langle W\rangle &\equiv& \frac{1}{\pi}\int_{-\pi/2}^{\pi/2} W(\rho_0,\xi)d\xi \nonumber \\
&\simeq& C\sqrt{\frac{2}{\pi}}\left[1-\frac{(\mu+5/4)}{2}\rho_0\right]\rho_0^{\mu+1/2}e^{-1/\rho_0}\, .
\end{eqnarray}

The choice of the optimal value for the normalised field amplitude $\rho_0=\Delta^2$ depends on several parameters, including the number of extracted electrons, the final needed beam quality, ion density, pulse peak electric field and size. If a large number of electrons has to be extracted, an optimal working point could be set so as the laser pulse is close to its saturation limit, \ie\, a large fraction of the ions in the vicinity of the pulse axis are  ionized after the pulse passage. The solution of Eq. \ref{eq:ADK} for a ionization depth $L$ is $n_e(L)=n_{i,0}(1-e^{-\bar{\Gamma}(L)})$ with 
\be
\label{eq:bargamma}
\bar{\Gamma}(L)=\int_0^L dz\langle W\rangle/c\, .
\ee
Setting $\bar{\Gamma}=1$ we get a ionization percentage of $\approx 60\%$, therefore $\bar{\Gamma}(L)\approx 1$ can be used to define the threshold of saturation effects. It is worth to define the local {\it average} spatial rate $\langle W\rangle /c$ as $\langle W\rangle /c\equiv \bar{k}_{ADK}\rho_0^{\mu+1/2} e^{-1/\rho_0}$, where
\be
\label{eq:bark}
\bar{k}_{ADK}=\sqrt{\frac{2}{\pi}} C(|m|)/c\, .
\ee
We are now able to find the normalised field bringing to a ($\approx 60\%$) saturation in a longitudinal length $L$. For the selected processes of Kr$^{8^+\rightarrow9^+} (m=0)$,
Ar$^{8^+\rightarrow9^+} (m=0)$, and
N$^{5^+\rightarrow6^+} (m=0)$, the $\bar{k}_{ADK}$ parameters evaluated with Eq. 2 in \cite{schroeder2014thermal} are $1.8\cdot 10^{5} \mu m^{-1}$, $1.4\cdot 10^{5} \mu m^{-1}$, $0.24\cdot 10^{5} \mu m^{-1}$, respectively. For each ionization process and saturation length $L$,  normalised field $\rho_0=a_0/a_c$ reaching saturation can be obtained by numerical solution of the equation
\be
\label{eq:satpar}
(\bar{k}_{ADK}L) \rho_0^{\mu+1/2} e^{-1/\rho_0}=1\, .
\ee
Graphical solutions of Eq. \ref{eq:satpar} for either tens-of-fs long pulses or near single-cycle pulses can be found in Appendix.

\section{Accurate residual momentum theory for a single cycle lasting ionization}
In this section we recast the theory for $\sigma_{u_x}$ and improve its accuracy  by i) including a ${\cal O} (\Delta^2)$ term not taken into account in \cite{schroeder2014thermal}, ii)  extending the theory up to ${\cal O} (\Delta^4)$ terms and, finally, iii)  including (exponential) correction terms due to the onset of saturation effects. We will start with local properties of the emitted electrons by neglecting saturation effects. Afterwards, we include the onset of saturation contribution for $\sigma_{u_x}$. The new analytical results can therefore be included in envelope codes aiming at an accurate statistical reconstruction of the ionization process even at ionization pulse intensities close to the {\it single cycle} saturation threshold (see below).     

\subsection{Local properties of the emitted electrons without saturation effects}
\begin{figure}[ht]
\centering
\includegraphics[width=\linewidth]{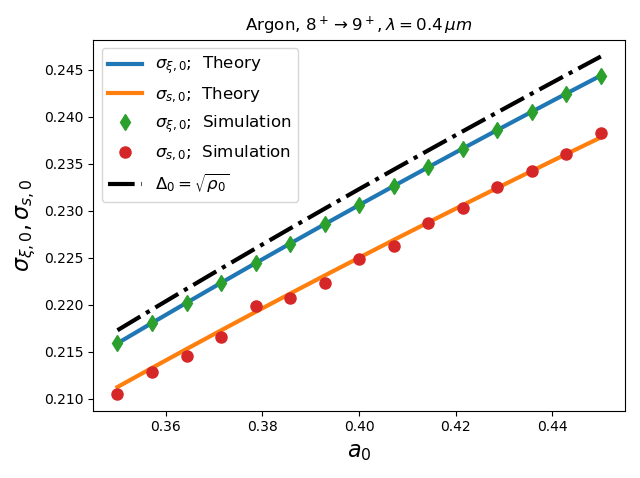}
\caption{Root mean square values of the local extraction phases $\xi_e$ and their sinus as a function of the laser amplitude $a_0$ ($\lambda_0 =0.4 \mu m$) for the process $Ar^{8^+\rightarrow 9^+}$. Blue curve shows the analytical results for $\sigma_{\xi,0}$ by Eq. \ref{eq:sigmaxis}, orange curve represents the analytical results for $\sigma_{s,0}$ by Eq. \ref{eq:sigmasinxis}. Results from Monte-Carlo simulations (green diamonds and red circles, respectively) agree well with theory. The black dash-dotted line refers to the bare (lowest order) estimation of $\sigma_{\xi,0}\simeq \sigma_{s,0}\simeq \Delta_0=\sqrt{\rho_0}$    }
\label{fig:Argon_xis}
\end{figure}

We start  considering  the \rms\, values of the extraction phase $\xi$ ($\sigma_\psi$ in \cite{schroeder2014thermal}) and of $\sin \xi$, with the aim of obtaining an approximated result including ${\cal O}(\Delta^4)$ (\ie\, ${\cal O}(\rho_0^2)$) corrections for the latter, but neglecting ionization saturation effects. 
Following \cite{schroeder2014thermal}, we consider a single half-cycle of the ionization pulse $E_x(\xi)=E_{0,x}\cos\xi$, extracting electrons with phases $\xi=k_0(z-ct)$ around the field  maximum at $\xi=0$. Expressing the ionization rate $W(\xi)$ in terms of the extraction phase, we get
\begin{eqnarray}
\label{Wxi}
W(\xi)&=&W_0\cdot (\cos\xi)^\mu\exp\left[\frac{1}{\rho_0}(\frac{1}{\cos\xi} -1)\right]\nonumber\\
&\simeq&W_0\exp\left[-\frac{\xi^2}{2\rho_0}\right]\left(1-\frac{\mu}{2}\xi^2-\frac{5}{24\rho_0}\xi^4\right) \nonumber\\
&\simeq&W_0\exp\left[-\frac{\xi^2}{2\sigma^2_\psi}\left(1+\frac{5}{12}\xi^2\right)\right]
\end{eqnarray}
where $W_0\equiv W(\xi=0)=k_{ADK}/k_0\rho_0^\mu e^{-1/\rho_0}$ is the maximum rate for the given pulse strength, $\sigma_\psi^{-2}=\rho_0^{-1}(1+\mu \rho_0)$ is the same expression of Eq. 6 in \cite{schroeder2014thermal}. The expansion of the exponential factor in Eq. \ref{Wxi} in powers of $\xi$ is justified by the fact that $\rho_0=\Delta^2\ll 1$ in our regimes. Here, terms containing $\xi^4/\rho_0$ are retained as they are ${\cal O}(\Delta^2)$ and this is related to the difference of our results from the equivalent terms in \cite{schroeder2014thermal} (see below). From now on, we will use $W(\xi)$ in the form $W_0e^{-\frac{\xi^2}{2\rho_0}}\left(1-\frac{\mu}{2}\xi^2-\frac{5}{24\rho_0}\xi^4\right)$, which results to be corrected up to ${\cal O}(\rho_0^2)$.   

It is now straightforward to evaluate the expectation values of $\xi^2$, obtaining (up to ${\cal O} (\Delta^2)$)
\be
\label{eq:sigmaxis}
\sigma^2_{\xi,0}\equiv\langle\xi^2\rangle=\rho_0\left(1-(\mu+5/2)\rho_0\right)\, ,
\ee
Our expression of $\langle\xi^2\rangle$ differs from the result in \cite{schroeder2014thermal} by the presence of the additional  $(-5/2)\rho_0$ term.  
  
The \rms\, residual momentum  $u_x=-a_0\sin\xi$ is, however, directly related to the {\it sinus} of the extraction phase $\xi$. Including all the correction terms up to $\rho_0^2$ but neglecting ponderomotive force and saturation contributions, we get $\sigma^2_{u,0}\equiv \langle u_x^2 \rangle=a_0^2\sigma^2_{s,0}$, where
\be
\label{eq:sigmasinxis}
\sigma^2_{s,0}\equiv\langle\sin^2 \xi\rangle=\rho_0\left(1+s_I\cdot \rho_0+s_{II}\cdot\rho_0^2\right)\, ,
\ee
$s_I=-(\mu+5/2+1)$ and $s_{II}=\frac{1}{8}(8\mu^2+68\mu+131)$. Once again, our expression up to the correction ${\cal O}(\rho_0)$ differs by the equivalent in \cite{schroeder2014thermal} by the presence of the $(-5/2)\rho_0$ term. Figure \ref{fig:Argon_xis} shows the dependence of $\sigma_{\xi,0}$ and $\sigma_{s,0}$ on the pulse amplitude $a_0$ for the local extraction of particles by the process $Ar^{8^+\rightarrow 9^+}$ and a pulse with wavelength $\lambda_0=0.4\, \mu m$.  For both the central moments the theory is able to reproduce the Monte Carlo simulations results with large accuracy.

\subsection{Local, single channel, ionisation process including  saturation effects}
Local saturation effects may be important when they occur within a {\it single} pulse cycle (see Fig. \ref{fig:saturation}). In this case, due to the monotonic reduction of the available ions as the pulse proceeds crossing each field peak, an asymmetry of the extraction average phase occurs, thus inducing a deviation of the \rms\, value for $u_x$ (see below) from the unsaturated case and the occurrence of a {\it nonzero} average momentum along the polarization axis. In this subsection we explore the local ionisation process occurring in a {\it single channel}, (e.g $Ar^{8^+\rightarrow 9^+}$), while multiple ionisation processes activated by the very large electric field will be discussed in the next subsection.

Going in deeper details with the rate equation \ref{eq:ADK}, we start expressing  the integral $\int (dn_e/dt) dt$ as   
\begin{eqnarray}
\label{eq:satweight}
\Gamma(\xi)&\equiv&\frac{1}{k_{0,x}}\int_{-\pi/2}^\xi dx W(x) \nonumber \\
&=&\frac{k_{ADK}}{k_{0,x}}\rho_0^\mu\int_{-\pi/2}^\xi dx (\cos x)^\mu e^{-\frac{1}{\rho_0 \cos x}} \nonumber\\
&\simeq& \nu_s(\rho_0){\cal G}\left(\frac{\xi}{\sqrt{2\rho_0}}\right)\, ,
\end{eqnarray}
where
\be
\label{eq:satshape}
{\cal G}(x)\equiv \frac{1}{2}(1+E(x))+\frac{\rho_0}{24\sqrt \pi}x(15+12\mu+10x^2)e^{-x^2}
\ee
is the saturation shape function, $E(x)$ is the error function, $k_{ADK}= C(|m|)/c$ and $\rho_0\ll 1$ has been used in the last manipulation. In Eq. \ref{eq:satweight} we have also introduced the saturation parameter $\nu_s={\bar \Gamma}(\lambda_x/2)$ (see Eqs. \ref{Wmean} and \ref{eq:bargamma}):
\be
\label{eq:nus}
\nu_s\equiv \sqrt{2\pi}\frac{k_{ADK}}{k_{0,x}}\left[1-\frac{(\mu+5/4)}{2}\rho_0\right]\rho_0^{\mu+1/2} e^{-\frac{1}{\rho_0}}\, .
\ee
\begin{figure}[ht]
\centering
\includegraphics[width=\linewidth]{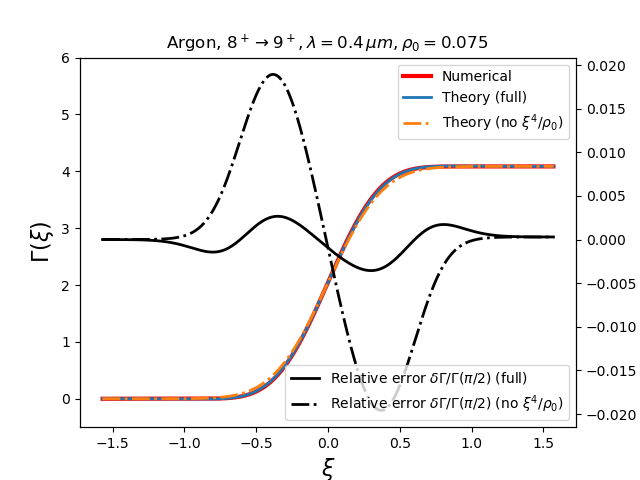}
\caption{Cumulative ionisation fraction $\Gamma(\xi)$ (see Eq. \ref{eq:satweight} evaluated numerically from the exact weight (red curve), from theory (blue curve) and by theory without the $\xi^4/\rho_0$ term (orange full-dashed line). The right axis shows the errors associated either with the  theory (black curve) or with the lower order theory without the non-gaussian $e^{-5\xi^4/(24 \rho_0)}$ correction.  }
\label{fig:Gamma}
\end{figure}

The saturation shape function ${\cal G}\left(\frac{\xi}{\sqrt{2\rho_0}}\right)$ accurately describes the particles extraction as the phase proceeds from $-\pi/2$ to $\xi$ within a single half pulse cycle and satisfies ${\cal G}\left(\frac{-\pi/2}{\sqrt{2\rho_0}}\right)= 0$, ${\cal G}\left(\frac{\pi/2}{\sqrt{2\rho_0}}\right)= 1$ provided that $\rho_0\ll1$. As it is apparent in Fig. \ref{fig:Gamma}, the full expression for ${\cal G}\left(\frac{\xi}{\sqrt{2\rho_0}}\right)$ predicts the (numerically evaluated) exact values for $\Gamma(\xi)$ with errors ${\cal O}(\rho_0^2)$, while the more simple expression 
\be
\label{eq:satshape0}
{\cal G}_0(x)\equiv \frac{1}{2}(1+E(x)) 
\ee
is also an accurate predictor, but with expected errors ${\cal O}(\rho_0)$.

Once the cumulative ionisation function $\Gamma(\xi)$ has been obtained,  the newborn electron distribution function equation, including saturation effects, can be evaluated as
\be
\label{eq:exactsat}
\frac{1}{n_{0,i}}\frac{dn_e}{d\xi}=-\frac{\partial}{\partial_\xi}e^{-\Gamma(\xi)}\, ,
\ee
which can be accurately approximated as
\be
\label{eq:weightsaturation}
\frac{1}{n_{0,i}}\frac{dn_e}{d\xi}=W_0e^{-\frac{\xi^2}{2\rho_0}}\left(1-\frac{\mu}{2}\xi^2-\frac{5}{24\rho_0}\xi^4\right)
e^{-\nu_s {\cal G}\left(\frac{\xi}{\sqrt{2\rho_0}}\right)}   
\ee
if $\rho_0\ll 1$.

The statistical local weight of Eq.  \ref{eq:weightsaturation} is now employed (instead of $W$ for the unsaturated case) to catch the cycle saturation effects on the extracted electrons phase-space distribution. Being now the weight {\it asymmetric} on any peak, the average extraction phase in any peak is no more null. 
To start with, we immediately evaluate the number of extracted electrons in the {\it first half cycle} as $n_e/n_{0,i}=\left(1-e^{-\Gamma(\xi=\pi/2)}\right)\simeq \left(1-e^{-\nu_s}\right)$. The statistical distribution of the extraction phase can strongly deviate from a Gaussian one once $\nu_s\gtrsim 1$, as the extraction phase can be modeled with a probability $P(\xi)\backsim dn/d\xi$ by using Eq. \ref{eq:weightsaturation}. To simplify the model, it is useful to work with a randomly distributed variable $x\in [-x_{max},x_{max}]$ with $x_{max}=\pi/(\sqrt{8\rho_0})$ and probability
\be
\label{eq:model}
P(x)\backsim  \left[1-\rho_0\left(\mu x^2+\frac{5}{6} x^4\right)\right]e^{-x^2-\nu_s {\cal G}(x)}\, ,
\ee
whose moments $\Xi(n,\rho_0)\equiv\langle x^n\rangle$  can be numerically evaluated as 
\be
\label{fig:csis}
\Xi(n,\rho_0)=\frac{\int_{-x_{max}}^{x_{max}} dx\, x^n P(x) }{\int_{-x_{max}}^{x_{max}} dx P(x) }\, .
\ee
The estimate of the average extraction phase within the peak  $\langle \xi_e\rangle_{single}$ reads now:
\be
\label{eq:singleum}
 \langle \xi_e\rangle_{single} \simeq \pm \sqrt{2\rho_0}\times \Xi(1,\rho_0)\,
\ee
where the sign of $\langle \xi_e\rangle_{single}$ depends of the phase of the field peak.  The second moment of the extraction phases can be evaluated in a similar way, obtaining
\be
\label{eq:singleu2m}
\langle \xi_e^2\rangle_{single} \simeq  2\rho_0 \times\Xi(2,\rho_0)\, .
\ee
 
\begin{figure}[ht]
\centering
\includegraphics[width=\linewidth]{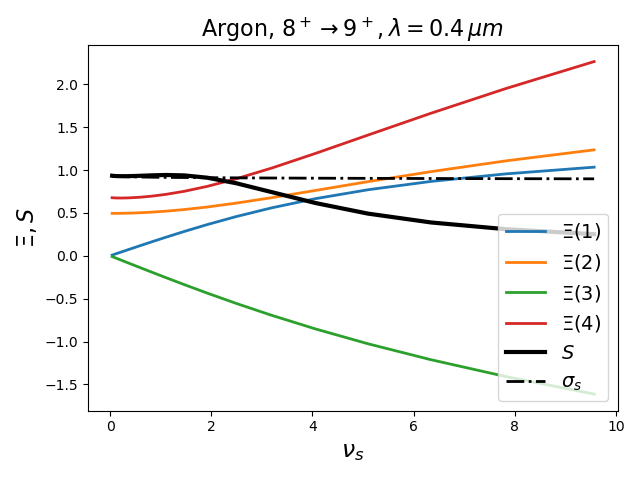}
\caption{Statistical moments $\Xi(n,\rho_0)$ for $n=1-4$   and full saturation correction $S$ numerically evaluated as in  Eqq. \ref{fig:csis},  and \ref{eq:Sfull}, as a function of the saturation parameter $\nu_s$ for the transition $Ar^{8^+\rightarrow 9^+}$ and $\lambda_0 =0.4\, \mu m$. 
}
\label{fig:CsiS}
\end{figure}
The moments $\Xi(n,\rho_0)$ for $n=1-4$, as a function of the saturation parameter $\nu_s$ and the ionisation process $Ar^{8^+\rightarrow 9^+}$ with $\lambda_0 =0.4\, \mu m$   are shown in Fig. \ref{fig:CsiS}. 
As a final result, in the case of partial or full saturation, the {\it single peak} distribution of the extraction phases around the local field maximum follows a {\it strongly non Gaussian} distribution of the shape as Eq. \ref{eq:model}, with $x=\xi_e/{\sqrt 2\rho_0}$ and an ionisation fraction of $(1-e^{-\nu_s})$. The resulting first and second order moments of the extraction phases follow Eqs. \ref{eq:singleum} and \ref{eq:singleu2m}.

Once the extraction phases have been statistically described, the resulting distribution of the residual transverse momenta is finally obtained (once again after neglecting ponderomotive force effects) by evaluating the particle momenta as $u_e=-a_0\sin(\xi_e)$. As the first peak ionises a fraction of the $(1-e^{-\nu_s})$ available ions, the remaining $e^{-\nu_s}(1-e^{-\nu_s})$ are extracted by the second peak of the cycle. There, as $\sin(\xi_e)$ changes its sign, a  {\it reversed} distribution of the momenta with respect to the first peak is obtained.  

It is interesting to note that a slight asymmetry and therefore a visible deviation from a Gaussian distribution occurs even at pulse amplitudes corresponding to (or close to) working points used in High-Quality beam production simulations (see e.g. \cite{tomassini2019high1}). 
\begin{figure}[ht]
\centering
\includegraphics[width=\linewidth]{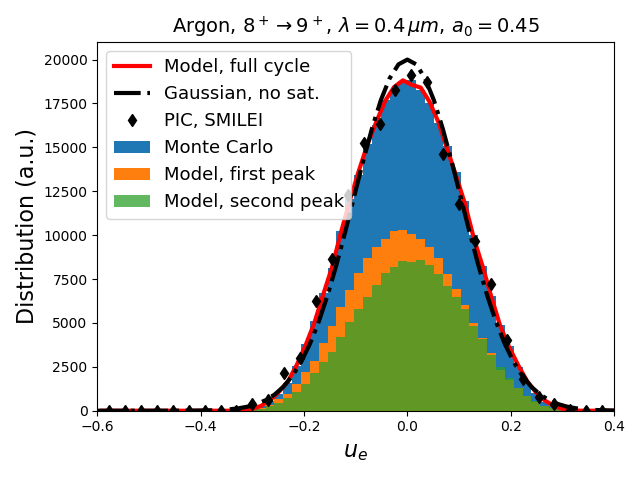}
\caption{Distribution of $u_e$ for the electrons extracted in a single cycle from Argon $8^+\rightarrow 9^+$ ions ($a_0=0.45$, $\lambda_0=0.4$ $\mu m$ corresponding to $\nu_s=0.252$). The blue bars show the distribution obtained  by a Monte-Carlo simulation. Orange and green bars refer to the distribution obtained  in the first and second peak, respectively, inferred by the model of Eq. \ref{eq:model}. 
}
\label{fig:Ar045}
\end{figure}
This is apparent in Fig. \ref{fig:Ar045}, where both the single peaks contributions from the model, as well as the full-cycle Monte Carlo and PIC \Smilei\thinspace simulations are shown together with the inferred Gaussian distribution obtained by using the \rms\, momentum as in Eq. \ref{eq:sigmasinxis}. There, the fraction $1-e^{-\nu_s}\simeq 22.3\%$ of the available ions are further ionised by the first peak and $e^{-\nu_s}(1-e^{-\nu_s})\simeq 17\%$ are extracted by the second peak. 
As a result, the model very accurately describes the process as it matches both the Monte Carlo and PIC simulations, while the standard Gaussian distribution partially deviates from the other distributions. 
\begin{figure}[ht]
\centering
\includegraphics[width=\linewidth]{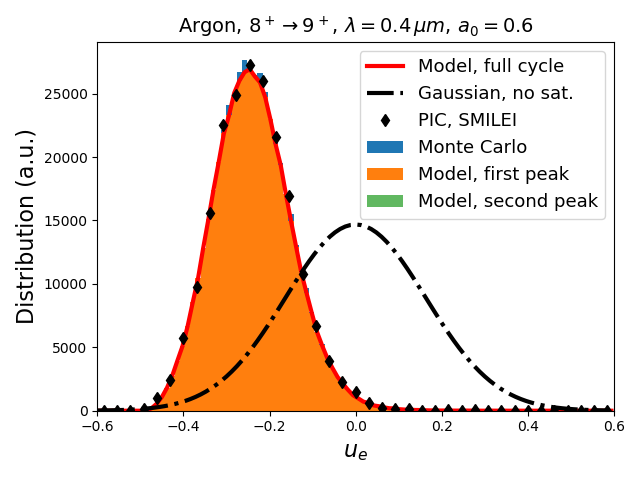}
\caption{Deep saturation distribution of $u_e$ for the electrons extracted in a single cycle from  $Ar^{8^+\rightarrow 9^+}$ processes ($a_0=0.6$, $\lambda_0=0.4 \mu m$ corresponding to $\nu_s=9.52$). 
Orange bars refer to the distribution obtained with the model of Eq. \ref{eq:model} (first peak of the cycle where more than $99.99\%$ of the available ions have been ionised). The blue bars are perfectly superimposed with the orange bars and show the distribution obtained  by a Monte-Carlo simulation. The green bars (not visible here due to the very few particles extracted there) show the distribution of the electrons extracted by the second peak of the cycle.  The red line  refers to the full-cycle electron distribution obtained by simulations {\it without} saturations effects, for reference.  
}
\label{fig:SatSingle}
\end{figure}
Moving into the deep-saturation territory, very large deviations from the standard Gaussian distribution are observed. 
Figure \ref{fig:SatSingle} compares the momenta distribution of the extracted electrons extracted  in the  case of deep saturation ($\nu_s=9.52\gg1$) for the  $Ar^{8^+\rightarrow 9^+}$ process ($a_0=0.6$, $\lambda_0=0.4 \mu m$). After the half pulse passage, about $99.998\%$ of the ions have been ionised

The analytical estimation of the cycle averaged mean and \rms\, momentum $u_x$ including saturation effects proceeds by observing that the cycle averaged sinus of the extraction phase can be evaluated averaging the contributes of the two peaks as
\be
\label{eq:cycleeum}
\langle \xi_e\rangle_{cycle}\simeq \sqrt{2\rho_0}\left[ \Xi(1,\rho_0)-\frac{1}{3}\rho_0\Xi(3,\rho_0)\right] \left(\frac{1-e^{-\nu_s}}{1+e^{-\nu_s}}\right)   \, \,
\ee
where Eq. \ref{eq:singleum} has been used. As the second phase moments of the two peaks in the cycle are exactly the same,  the cycle averaged $\langle \xi_e^2\rangle_{cycle}$ can be evaluated directly from Eq. \ref{eq:singleu2m}. As a result, the  full cycle averaged  central momentum of the electron locally extracted by a single ionisation process is evaluated as  $\sigma_{u_x}\equiv <u_x^2>-<u_x>^2=a_0^2\sigma^2_s$, where 
\be
\label{eq:sigmausat}
\sigma_s^2\simeq\sigma_{s,0}^2\, S(\nu_s)
\ee
and the overall saturation correction $S(\nu_s)$ is 
\begin{eqnarray}
\label{eq:Sfull}
S(\nu_s)&\equiv& 2\Xi(2,\rho_0)-\frac{4}{3}\rho_0\Xi(4,\rho_0)+ \nonumber\\
&-&2\left\{
\left[\Xi(1,\rho_0)-\frac{1}{3}\rho_0\Xi(3,\rho_0)\right]
\frac{1-e^{-\nu_s}}{1+e^{-\nu_s}}\right\}^2\, .  
\end{eqnarray}
The overall saturation correction slightly increases above unity in the range $0\lesssim \nu_s\lesssim 1$ (see the black line in Fig. \ref{fig:CsiS}). In this range, both the peaks in each pulse contribute in extracting particles with opposite average momenta, thus inducing an increase of the \rms\, full cycle transverse momentum. In the deep saturation regime ($\nu_s\gtrsim 1$) the second peak gives even more negligible  contribution while the single peak \rms\, momenta  decrease due to the phase space cut induced by the strong saturation, with the final result of generating an overall \rms\, momentum well below the one expected without saturation effects on. 
\begin{figure}[ht]
\centering
\includegraphics[width=\linewidth]{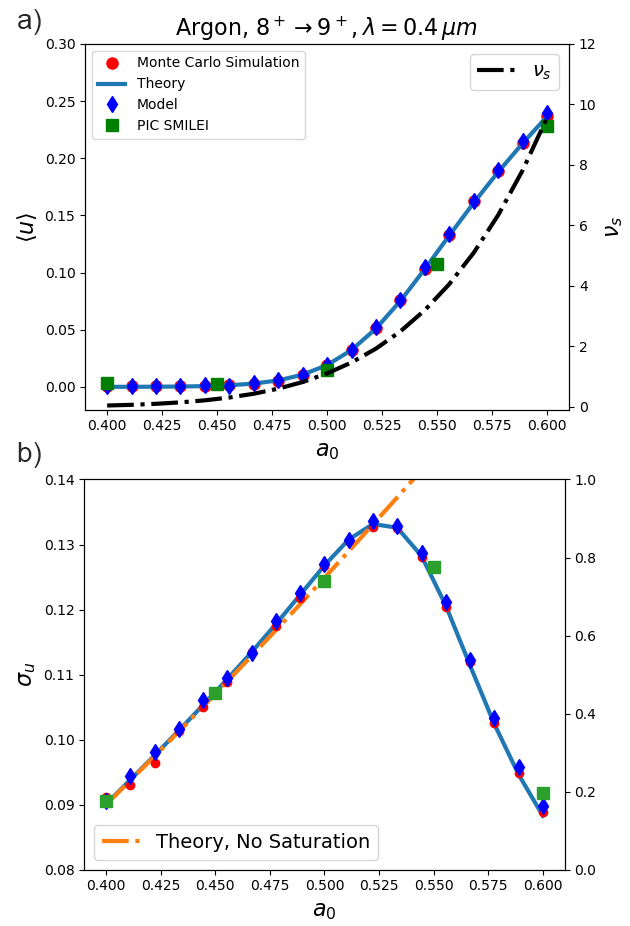}
\caption{Average and \rms\, residual momentum for the channel Argon$8^+\rightarrow 9^+$, single pulse cycle with $\lambda_0=0.4 \mu m$, as a function of the pulse amplitude $a_0$. a) Average momentum as expected by theory (blue line), by Monte-Carlo simulations (red circles), by using the model of Eq. \ref{eq:model} (blue triangles), and by \Smilei\thinspace PIC simulations (green squares). The black right axis refers to the ionisation fraction after one pulse cycle. b) Root mean square of the residual momenta. The blue line shows the analytical results which include the saturation effects through the $S(\nu_s)$ function. The orange full-dashed line shows the analytical results without saturation effects, for reference. Red circles, blue triangles and green squares show the results by Monte-Carlo, by the model and by \Smilei\thinspace PIC simulations, respectively.  
}
\label{fig:Argonsinglecycle}
\end{figure}
The final results for the cycle averaged first and second order moments of the residual momenta in the case of the single process $Ar^{8^+\rightarrow 9^+}$ are shown in Fig. \ref{fig:Argonsinglecycle}. As we clearly see in Fig. \ref{fig:Argonsinglecycle}-b), if $\lambda_0 =0.4\, \mu m$ the maximum \rms \, momentum is achieved with $a_0\approx 0.53$. We stress that those results are obtained by activating the single ionisation channel described above.

\subsection{Single-cycle, multiple channel ionisation processes}
\begin{figure}[ht!]
\centering
\includegraphics[width=\linewidth]{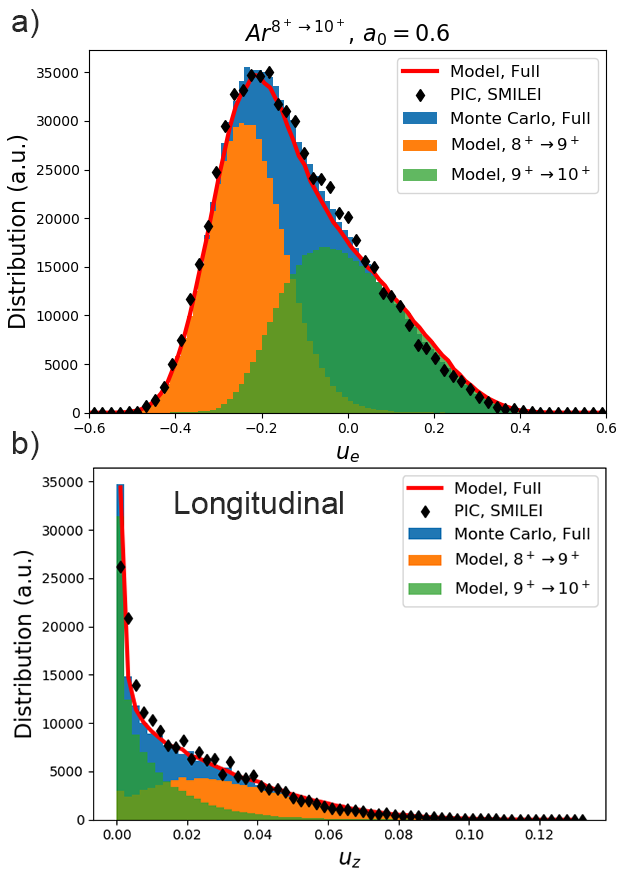}
\caption{ 3D distribution of the  residual momentum for the $(0)$ and $(1)$ channels $Ar^{8^+\rightarrow 9^+}$ and $Ar^{9^+\rightarrow 10^+}$ in the deep saturation regime,  single pulse cycle with $a_0=0.6$ and $\lambda_0=0.4\, \mu m$.
The blue bars and the black curve show the distribution of the full process
$Ar^{8^+\rightarrow 10^+}$ as inferred by a Monte Carlo simulation and by \Smilei\thinspace PIC simulations, respectively. Orange and green bars show the distribution obtained by the model for the channels $Ar^{8^+\rightarrow 9^+}$ and $Ar^{9^+\rightarrow 10^+}$, respectively. Panel a) depicts the residual transverse momentum distribution along the polarisation axis $x$, while in panel b) the longitudinal residual momentum $u_z$ is shown. Since ponderomotive forces are not taken into account, the residual momentum along $y$ is zero (not shown here). As it is clear from the sum of the $(0,1)$ channels (red line), the model is capable to well reproduce the single-cycle momenta distribution even in a multi-channel regime.  
}
 \label{fig:Argon2channels}
\end{figure}
In the single-cycle intermediate and deep saturation regimes, the pulse electric field is usually large enough to activate one (or more) ionisation channel(s) above the starting, selected one.  Referring to the usual Argon example,  when $\nu_s\gtrsim 1$ a two-channels process related to the ($l=1$, $m=0$)  $Ar^{8^+\rightarrow 9^+}$, $Ar^{9^+\rightarrow 10^+}$  occurs, with the next process $Ar^{10^+\rightarrow 11^+}$ ($m=1$) having a statistical weight significantly lower than the others. The analysis reported in the previous subsection can be applied on the single channels, thus giving insight into the whole ionisation process. To start with, we denote with the subscripts $(0)$  and $(1)$ the base (selected) process and the subsequent one, respectively, and with $n^{(0)}_i$ $n^{(1)}_i$ their initial available ions. 

The total number of extracted electrons in any peak can be obtained by solving the rate equations for the local available ions 
\begin{eqnarray}
\label{eq:rateeq_twolevels}
 \begin{cases} \frac{dn^{(0)}}{d\xi} = -n^{(0)} \nu_s^{(0)} {\cal G}^{(0)} \\ 
\frac{dn^{(1)}}{d\xi} = -n^{(1)} \nu_s^{(1)} {\cal G}^{(1)}+n^{(0)} \nu_s^{(0)} {\cal G}^{(0)}  \end{cases}
\end{eqnarray}
whose solutions give the total number of extracted electrons in any process and their distribution. 
As shown before, the number of electrons extracted in any peak by the processes $(0)$ and $(1)$ are
\begin{eqnarray}
\label{eq:extracedelectrons}
N_e^{(0)}&=&n^{(0)}_i(1-e^{-\nu_s^{(0)}})\nonumber\\
N_e^{(1)}&=&n^{(1)}_i(1-e^{-\nu_s^{(1)}})+\nonumber\\
&+&n^{(0)}_i\left(1-e^{-\nu_s^{(0)}}-e^{-\nu_s^{(1)}}{\cal M}_{01}\right)\, ,
\end{eqnarray}
where the transfer function  ${\cal M}_{01}(\rho_0;\xi)$ is defined as 
\be
{\cal M}_{01}(\rho_0;\xi)\equiv W_0^{(0)}\int_{-\pi/2}^\xi dt e^{\nu_s^{(1)} {\cal G}^{(1)}(t)} P^{(0)}(t)
\ee
\begin{figure}[ht]
\centering
\includegraphics[scale=0.5]{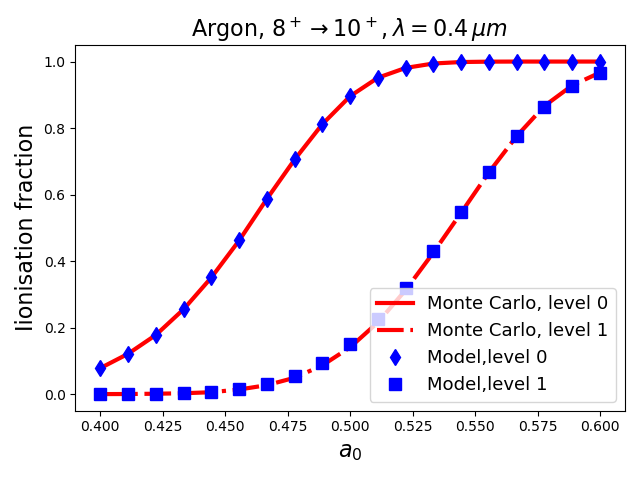}
\caption{Ionisation fraction in the channels $(0)$ and $(1)$ as a function of the pulse amplitude for the case $Ar^{8^+\rightarrow 10^+}$, $\lambda_0 =0.4\, \mu m$. The red lines refer to the predictions from Eq. \ref{eq:extracedelectrons}, while the blue points are obtained by Monte Carlo simulations. Predictions with errors ${\cal O}(\rho_0^2)<1\%$ are obtained in this way.  
}
\label{fig:SatFraction}
\end{figure}
Equations \ref{eq:extracedelectrons} very accurately predict the number of extracted electrons in any channel in a a single pulse peak, being the maximum discrepancy between the inferred number of extracted electrons and Monte Carlo simulations outcomes below $1\%\approx\rho_0^2$ (see Fig. \ref{fig:SatFraction})

The distribution of the extracted electrons in the channel $(0)$ follows the already discussed prescriptions from Eq. \ref{eq:model}. The distribution from process $(1)$ takes origin both from the ions initially available at level $(1)$ and those that are freed while the phase proceeds within the peak. As the exact expression of the distribution
\be
\label{eq:secondleveldistribution}
\frac{dn_e^{(1)}}{d\xi}=W_0^{(1)}P^{(1)} 
\left[n_i^{(1)}+n_i^{(0)}{\cal M}_{01}(\rho_0;\xi)
\right]
\ee
contains the transfer function ${\cal M}_{01}(\rho_0;\xi)$ that would be evaluated numerically for any $\xi$, we just evaluate ${\cal M}_{01}(\rho_0;\pi/2)$ so as to accurately infer the number of extracted electrons, whose  distribution is modelled by making the approximation
\be
\int_{-\pi/2}^\xi dt e^{\nu_s^{(1)} {\cal G}^{(1)}(t)} P^{(0)}(t)\simeq e^{\nu_s^{(1)} {\cal G}^{(1)}(\xi)}\left(1-e^{\nu_s^{(0)} {\cal G}^{(0)}(\xi)}\right)\, .
\ee
The approximation is accurate because not negligible values for  $\nu_s^{(1)}$ are necessary linked to a saturated  regime of the base level, which realises quasi-flat injection of available ions of the second level. 
\begin{figure}[ht]
\centering
\includegraphics[width=\linewidth]{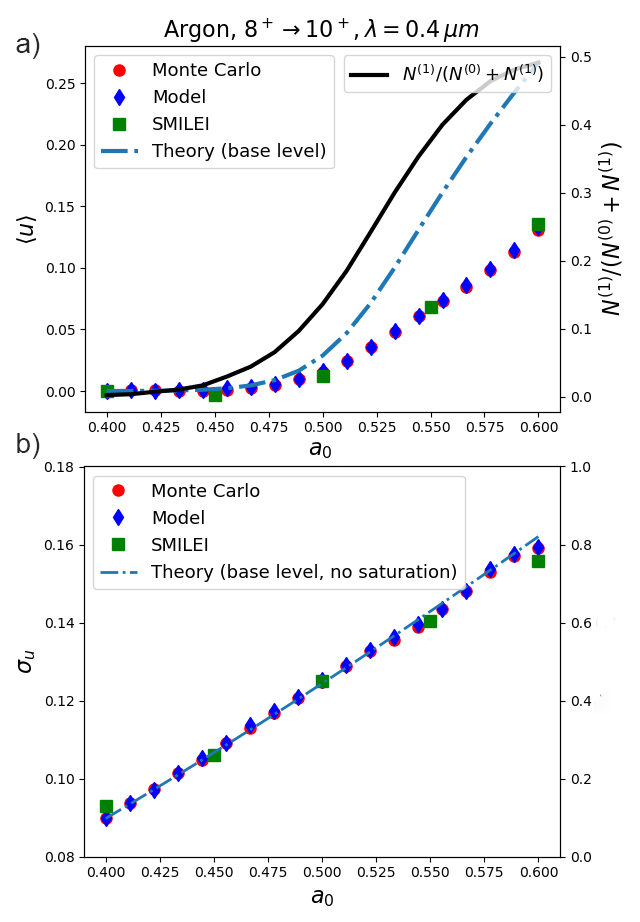}
\caption{Single cycle, two-levels ionisation scan for the $Ar^{8^+\rightarrow 10^+}$ process with $\lambda_0 =0.4\, \mu m$.  Red circles, blue diamonds and green square refer to Monte Carlo simulations, model predictions and PIC simulations, respectively. a) Average momentum from the two-levels simulations and the model, as well as the average momentum as predicted by the {\it single} base level $Ar^{8^+\rightarrow ^9+}$, for reference (blue line). The vertical axis on the right shows the fraction of level $(1)$ over the the whole $(0)+(1)$ particles extracted in the cycle. b) \rms\,  momentum from the two-levels simulations and the model. The blue line shows predictions by the theory of the base level {\it without} saturation effects on.   
}
\label{fig:Artwolevels}
\end{figure}

The two-levels model for the whole process occurring in a single peak, including the estimates of the extracted particles {\it via} Eq. \ref{eq:extracedelectrons} and extraction phase distributions following the base level distribution Eq.  \ref{eq:model}  and \ref{eq:secondleveldistribution}, can be combined so as to get the whole $(0)+(1)$ process (\eg\, $Ar^{8^+\rightarrow 10^+}$) in a full pulse cycle. Figure \ref{fig:Artwolevels} shows the {\it full-cycle} scan of the average and \rms\, momentum for the two-levels process $Ar^{8^+\rightarrow 10^+}$ with $\lambda_0=0.4\, \mu m$, as a function of pulse amplitude $a_0$. The model prediction (blue diamonds) agree with Monte Carlo simulations (red circles) and PIC simulations (green squares) for both the average momentum (box a) and for the \rms \, momentum (box b). The black line from the right axis in a) shows the fraction of the second ionisation process $(1)$ over the whole set of particles extracted in the cycle, showing that the model maintains its accuracy also in the case of the second ionisation deep saturation. The model can be easily extended in order to include relevant contribution of further ionisation processes. 

In Fig. \ref{fig:Artwolevels}-a), a blue line representing the average momentum as predicted by the {\it single} $Ar^{8^+\rightarrow 9^+}$ process shows that the second ionisation step induces a sensible reduction of the average momentum as a large ionisation of the $(0)$ level in the first first peak causes an increase of the number of particles extracted in level $(1)$ during the second peak, where $\sin(\xi_e)$ has an opposite sign.
Furthermore, in Fig. \ref{fig:Artwolevels}-b) we can also note that the additional $(1)$ level rules out the momentum drop off induced by saturation in  the single $(0)$ process. As  a matter of fact, both the model and the simulations outcomes fit surprisingly well with predictions by the theory of {\it unsaturated} ionisation by the single $(0)$ process (see also the blue line in b), representing results from Eq. \ref{eq:sigmasinxis}.

\section{Whole bunch emittance theory}
As the cycle pulse amplitude depends on both the longitudinal and transverse coordinate, we make the substitution $a_0\rightarrow f(\vec{x})a_0$, being $f$ the laser pulse envelope shape. As a result, for any position $\vec x$ the statistical average weight of the extracted electrons Eq. \ref{Wmean}, as well as their \rms\, transverse momentum, depend on $\vec{x}$ trough $f$.   We move on by firstly neglecting saturation effects (see Fig. \ref{fig:saturation}) and ponderomotive force effects.

\subsection{Theory with negligible saturation effects}
The description of the spatial dependence of $\sigma_{u_x}$. and subsequent evaluation of the whole beam emittance, can be simplified by introducing the generating functional of the spatial moments: 
\begin{eqnarray}
\label{eq:generatingmoments}
{\cal G}(m,n)&\equiv& \langle e^{-mr^2-n(z-ct)^2}\rangle \nonumber\\
&=&\frac{\int d^3x e^{-mr^2-n(z-ct)^2} dn_e/dt({\vec x})}{\int d^3x dn_e/dt({\vec x})}\, ,
\end{eqnarray}
where $dn_e/cdt = \langle W\rangle$ has been used in absence of saturation effects (see Eq. \ref{Wmean}) and $\rho=\rho_0 f$ includes the pulse envelope $f$ effects. If the pulse envelope gas a bi-gaussian shape $f(r,z-ct)=\exp(-r^2/w_0^2-(z-ct)^2/L^2)$,  the transverse functional ${\cal G}(k,0)=\langle e^{-k\frac{r^2}{w_0^2}}\rangle$ is evaluated {\it without further approximations} by means of integrals of the form 
\begin{eqnarray}
\label{eq:I}
I(k,\rho_0)&\equiv&\int_0^\infty dx^2 e^{\left[-(\mu +\frac{1}{2}+k)x-\frac{1}{\rho_0}(e^{x^2}-1)\right]} \nonumber\\
&=&e^{-(\mu +\frac{1}{2}+k)+\frac{1}{\rho_0}}
\Gamma^{up}\left[-(\mu+\frac{1}{2}+k);\frac{1}{\rho_0}\right]
\end{eqnarray}
being $\Gamma^{up}(s,x)$ the upper incomplete Euler function $\Gamma^{up}(s,x)=\int_x^\infty dt e^{-t} t^{s-1}$. As a result, we get:
\begin{eqnarray}
\label{eq:Gperp}
{\cal G}(k,0)&\equiv& \langle e^{-k r^2/w_0^2}\rangle \nonumber \\
&=&\frac{I(k,\rho_0)-(\frac{\mu}{2}+\frac{5}{8})\rho_0 I(k+1,\rho_0) }{I(0,\rho_0)-(\frac{\mu}{2}+\frac{5}{8})\rho_0 I(1,\rho_0)} \nonumber \\
&\simeq& 1-k \rho_0+k(\mu+\frac{5}{2})\rho_0^2+{\cal O}(\rho_0)^3
\end{eqnarray}
We stress that, depending upon the needed accuracy, it is possible to use either the expression containing the Euler incomplete Gamma functions or its (less accurate) polinomial expansion.  

The longitudinal counterpart of Eq. \ref{eq:Gperp}, \ie\,  ${\cal G}(0,k)\equiv \langle e^{-k (z-ct)^2/L^2}\rangle$, can be evaluate in a similar way. We observe, however, that for any $k\in \Re$ we get
\be
\langle e^{-k x^2/w_0^2}\rangle=\langle e^{-k y^2/w_0^2}\rangle=\langle e^{-k (z-ct)^2/L^2}\rangle
\ee
which brings to 
\begin{eqnarray}
\label{eq:Glong}
{\cal G}(0,k)&=&\sqrt{{\cal G}(k,0)}\nonumber\\
&\simeq& 1-\frac{1}{2}k\rho_0+\frac{1}{2}k\left[(\mu+\frac{1}{2})+\frac{3}{4}k\right]\rho_0^2\, .
\end{eqnarray}
The full average generator is finally evaluated as
\begin{eqnarray}
\label{eq:Gfull}
{\cal G}(k,k)&\equiv& \langle e^{-k(r/w_0)^2-k (z-ct)^2/L}\rangle=\left({\cal G}(k,0)\right)^{3/2}\nonumber\\
&\simeq& 1-\frac{3}{2}k\rho_0+\frac{3}{2}k\left[(\mu+\frac{1}{2})+\frac{5}{4}k\right]\rho_0^2\, .
\end{eqnarray}
The first usage of ${\cal G}(k,k)$  is for the evaluation of the {\it whole bunch} \rms\, value of the residual momentum $u_x$. This can be performed by observing that $\langle \rho^k\rangle=\rho_0^k{\cal G}(k,k)$, obtaining
\begin{eqnarray}
\label{eq:sigmaubunch}
\langle\sigma^2_u\rangle &\equiv& \frac{\int d^3x \sigma^2_{u_x}\times dn_e/dt({\vec x})}{\int d^3x dn_e/dt({\vec x})} \nonumber\\
&=&a_c^2\rho_0^3\left[
{\cal G}(3,3)+s_I\rho_0
{\cal G}(4,4)+s_{II}\rho_0^2{\cal G}(5,5)\right]\, .
\end{eqnarray}
We stress here that ${\cal G}(k,k)$ can be evaluated without further approximations by using the incomplete Euler Gamma functions in 
Eq. \ref{eq:I}. A faster evaluation of $\langle\sigma^2_{u_x}\rangle$, however, can be obtained by Taylor expanding Eq. \ref{eq:sigmaubunch}  with corrections up to ${\cal O}(\rho_0^2)$, obtaining
\begin{eqnarray}
\label{eq:sigmaubunchfinale0}
\sigma^2_{u_x,bunch,0}&\equiv& \langle\sigma^2_u\rangle_{bunch}\simeq a_0^2\rho_0\times \nonumber\\
&\times&\left[ 1-(\mu+8)\rho_0+(\mu^2+19\mu+\frac{131}{2})\rho_0^2 \right]\, .
\end{eqnarray}
The difference with the equivalent result in \cite{schroeder2014thermal} (see Eq. 14) is, as in the local analysis, twofold: our $\Delta^2=\rho_0$ correction term differs from the equivalent one in \cite{schroeder2014thermal} and we included a $\Delta^4=\rho_0^2$ contribution. The $\rho^2$ term in Eq. \ref{eq:sigmaubunchfinale0} is {\it not} a tiny contribution, as the prefactor $(\mu^2+19\mu+\frac{131}{2})$ ($\approx 15$ for the krypton, $\approx 30$ for the argon and $\approx 50$ for the nitrogen) is usually large. 
\begin{figure}[ht]
\centering
\includegraphics[scale=0.56]{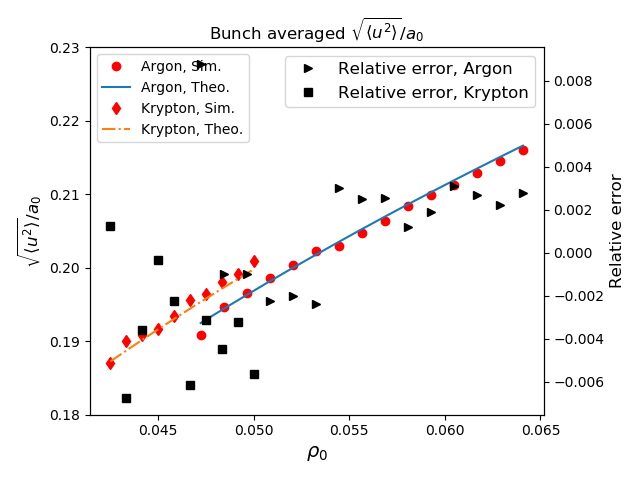}
\caption{Whole bunch \rms\, momentum as a function of the normalised field strength $\rho_0=a_0/a_c$ for a process without saturation and ponderomotive force effects. Diamond and circle points represent simulation results for krypton and argon, respectively. The orange and blue lines show, for the same processes, the analytical results from Eq. \ref{eq:sigmaubunchfinale0}. In the right axis, the relative errors committed by the analytical formulae are shown as  black points (squares for krypton and triangles for argon). In both cases, a relative error below $1\%$ is expected. 
}
\label{fig:wholebunchnosat}
\end{figure}
In Fig. \ref{fig:wholebunchnosat} the analytical results of Eq. \ref{eq:sigmaubunchfinale0} (dashed lines) are compared with simulations which exclude either saturation of the ionization process and ponderomotive force effects in the subsequent electron dynamics inside the laser field. In this case, errors below $1\%$ are expected when evaluating the full bunch \rms \, momentum along the laser polarisation axis.

The functional generator of the moments ${\cal G}(m,n)$ can be employed, of course, to evaluate the \rms \, values of the transverse and longitudinal bunch size, too. This can be accomplished by observing that, for any slice at fixed $z-ct$, the \rms\, extraction radius can be evaluated as
\be
\langle r^2\rangle=-\partial_m {\cal G}(m,0)_{m=0}\, .
\ee
The gradient $\partial_m {\cal G}(m,0)$ can be obtained either in an exact form by using the complete version of $I(k,\rho_0)$ as in Eq. \ref{eq:I}, or by referring to its polinomial expansion in $\rho_0\ll 1$. In the last case, we get (for a fixed slice $z-ct$):
\be
\langle r^2\rangle \simeq w_0^2 \rho_0 \left[1-(\mu +\frac{5}{2})\rho_0\right]
\ee
A further average over the longitudinal $z-ct$ slices will give us the {\it whole bunch} \rms\, transverse size
\begin{eqnarray}
\label{eq:rmsbunch0}
\sigma^2_{x,bunch,0}&\equiv &\langle x^2\rangle_{bunch} \simeq \frac{1}{ 2} w_0^2 \rho_0\times \nonumber\\
&\times&\left[1-(\mu +3)\rho_0+\frac{1}{2}(3\mu+\frac{33}{4})\rho_0^2\right]\, .
\end{eqnarray}
As a final result, as $\langle x u_x\rangle =0$, the whole beam normalised emittance squared along the polarisation axis (excluding saturation and ponderomotive effects) reads
\begin{eqnarray}
\label{eq:beamemittancenosat}
\epsilon^2_{n,x}&\equiv& \langle x^2\rangle_{beam}
\langle u_x^2\rangle_{beam}-\left( \langle x u_x\rangle_{beam}\right)^2 \nonumber\\
&=&\frac{1}{2}\left(a_0\, w_0\, \rho_0\right)^2 {\cal E}_n(\rho_0,\mu_0) \, , 
\end{eqnarray}
where the universal emittance correction  term ${\cal E}_n(\rho_0,\mu)$ can be evaluated retaining ${\cal O}(\rho^2)$ terms as
\be
\label{eq:beamemittancecorrection}
{\cal E}_n(\rho_0,\mu)\simeq 1-(\mu+11)\rho_0+\left(2\mu^2+\frac{63}{2}\mu+\frac{749}{8} \right)\rho_0^2\, .
\ee
Equations \ref{eq:beamemittancenosat} and \ref{eq:beamemittancecorrection} correctly describe the whole beam emittance in the case of negligible saturation, as it is apparent in Fig. \ref{fig:wholebunch}-c), where the orange line matches with simulations relative to low values of $\rho_0$. Furthermore, we also note that the model fits (with unsaturated working points)  with simulations {\it including} ponderomotive force effects, as those effects don't increase the beam emittance (at least at the leading order) \cite{schroeder2014thermal}.

\subsection{Whole bunch quality including saturation effects}

\begin{figure}[ht!]
\includegraphics[width=\linewidth]{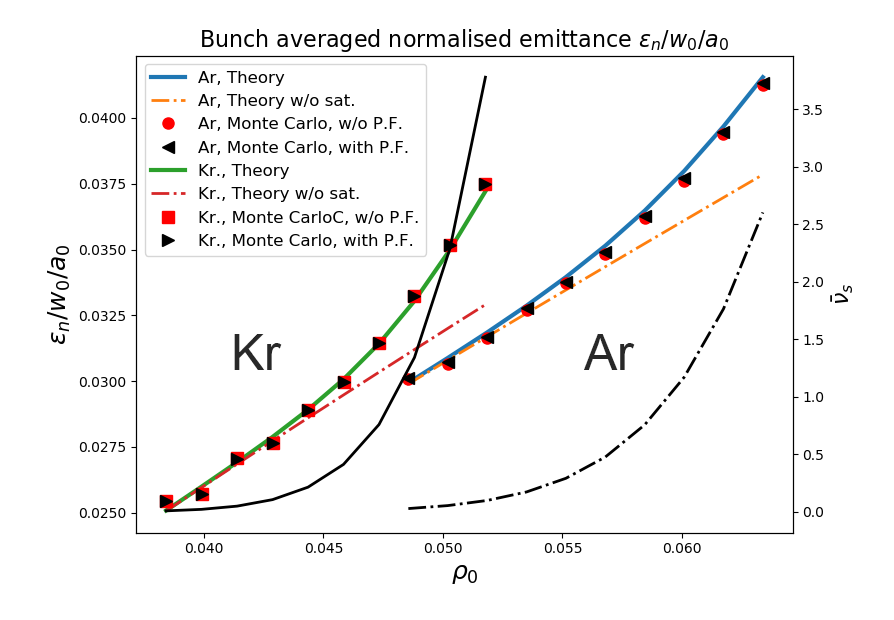}
\caption{ Bunch averaged normalised emittance obtained with a thin slice of ionisable atoms (either krypton or argon) with a scan on the normalised field strength $\rho_0=a_0/a_c$. The pulse wavelength, waist and durations are $0.4\mu m$, $5\mu m$ and $10fs$, respectively. The emittance is further normalised by the pulse waist $w_0$ and amplitude $a_0$, \ie\, $\epsilon_n/(w_0a_0)=\sqrt{\langle u^2\rangle\langle x^2\rangle-(\langle ux\rangle)^2}/(w_0a_0)=\rho_0\sqrt{{\cal E}_n}$. Here black points represent simulation results of  simulations {\it  including   } ponderomotive force effects, while red points refer to simulations withoutv ponderomotive force effects on.   Diamond and circle points represent simulation results for krypton and argon, respectively, which include saturation effects during  ionization but exclude the ponderomotive force contribution in the subsequent particles evolution. The dashed lines show, for the same processes, the analytical results with excludes saturation effects. Thick lines show the analytical results  with a full description of the ionization process. 
}
\label{fig:wholebunch}
\end{figure}
 
The onset of ionization saturation  during the {\it whole pulse passage} usually occurs at pulse amplitudes close to those selected as working points, \ie\, lower than those necessary to get saturation effects within a single pulse peak.  A first effect is the {\it reduction} of the number of particles extracted in the vicinity of the pulse axis, thus enhancing the statistical weight of the regions with $r\backsimeq \Delta_0 w_0$ and therefore increasing the final $\langle r^2\rangle_{bunch}$. As a result, the \rms\, residual momentum is slightly smaller than that expected without saturation effects on. We will see however that, as anticipated in \cite{schroeder2014thermal} the final effect is that of a whole-bunch emittance increase, being the final result dominated by the increase of the bunch radius, indeed.

The integrated ionisation weight $\bar{\Gamma}(r, z-ct)$ can be evaluated as ($\rho_0\ll 1$)
\begin{eqnarray}
{\bar{\Gamma}}(r, z-ct)&=&\int_{-\infty}^{z-ct}\langle W(r,\zeta)/c\rangle d\zeta  \nonumber\\
 &\simeq& \bar{\nu}_s e^{-\frac{r^2}{\rho_0w_r^2}}\times
 \frac{1}{2}\left[1+E\left(\frac{z-ct}{{\sqrt \rho_0}w_z} \right)   \right]\,
\end{eqnarray}
where 
\be
\label{barnu0}
\bar{\nu}_s={\sqrt 2}(k_{ADK}w_z)\rho_0^{\mu+1}e^{-1/\rho_0}\, .
\ee
We can now use $e^{-\bar\Gamma(r,z-ct)}\langle W\rangle(r,z-t)$ as a weight to obtain \rms\, quantities. Starting with the \rms \, beam radius $\langle r^2\rangle_{sat}$ we get at the lowest order in $\rho_0$:
\begin{eqnarray}
\label{eq:rmsbeamsat}
\langle r^2\rangle_{sat}&=&\frac{\int_{-\infty}^\infty dz \int_0^\infty dr^2 r^2 \langle W\rangle e^{-\bar{\Gamma}} }{\int_{-\infty}^\infty dz \int_0^\infty dr^2 \langle W\rangle e^{-\bar{\Gamma}}}\nonumber\\
&=&\frac{\int_0^\infty dr^2 r^2\left(1- e^{-\bar{\nu}_se^{-r^2/w_r^2} }\right)}{\int_0^\infty dr^2 \left(1- e^{-\bar{\nu}_se^{-r^2/w_r^2} }\right)}\nonumber\\
&\approx& w_0^2\rho_0\left(1+\frac{1}{8}\bar{\nu}_s-\frac{5}{864}\bar{\nu}_s^2 +{\cal O}(\bar{\nu}_s^3)\right)\,
\end{eqnarray}
where the last expression holds for $\bar{\nu}_s \ll 1$. 

The evaluation of the \rms\, momentum including saturation effects proceeds by generalising the generating functional of the moments ${\cal G}(m,n)$ (Eq. \ref{eq:generatingmoments}) so as to include the progressive decrease of the available ions as the the comoving coordinate $z-ct$  proceeds towards the tail of the pulse. Once again, we will get only the lowest order corrections in $\rho_0$ and $\bar{\nu}_s$, obtaining for the special case of interest
\begin{eqnarray}
\label{eq:gefunzsat}
{\cal G}(3,3)_{sat} &\simeq& \frac{\iint dx^2\,d\zeta  e^{-3(1+\frac{1}{\rho_0})(x^2+\zeta^2)-\frac{\bar{\nu}_s}{2}e^{-x^2}\left(1+E(\zeta)\right)} }{\iint dx^2\,d\zeta  e^{-\frac{3}{\rho_0}(x^2+\zeta^2)-\frac{\bar{\nu}_s}{2}e^{-x^2}\left(1+E(\zeta)\right)} }\nonumber\\
&\simeq& \left(1-\frac{9}{2}\rho_0\right)\left(1-\frac{3}{8}\rho_0\bar{ \nu}_s\right)\, ,
\end{eqnarray}
where in the last manipulation we retained the lowest order in $\bar{\nu}_s$ and used $\int_{-\infty}^{\infty}dxE(x)e^{-ax^2}=0$ for $a>0$.  
By inspection of Eqq. \ref{eq:Gfull} and \ref{eq:sigmaubunch}, we note that $\langle r^2\rangle_{sat}$ and ${\cal G}(3,3)_{sat}$  contain  corrections to their leading terms in $\rho_0$. Collecting the saturation corrections into the whole bunch normalised emittance, which now contains the leading order correction terms due to saturation effects, we get 
\be
\label{eq:beamemittancesisat}
\epsilon^2_{n,x}\simeq\frac{1}{2}\left(a_0\, w_0\, \rho_0\right)^2 {\cal E}_{n,sat}(\rho_0,\mu_0) \, , 
\ee
where the emittance correction  term ${\cal E}_{n,sat}(\rho_0,\mu)$ including  saturation effects with $\bar{\nu}_s\ll 1 $ is 
\begin{eqnarray}
\label{eq:beamemittancecorrection2}
{\cal E}_{n,sat}&\simeq& \left(1+\frac{\bar{\nu}_s}{8}-\frac{5}{864}\bar{\nu}_s^2\right)\times \nonumber\\
&\times&\left[1-(\mu+11+\frac{3}{8}\bar{\nu}_s)\rho_0+\right. \nonumber\\
&+& \left. \left(2\mu^2+\frac{63}{2}\mu+\frac{749}{8}+\frac{3}{8}(\mu+11)\bar{\nu}_s \right)\rho_0^2\right]\, .
\end{eqnarray}
Although the results from Eq. \ref{eq:beamemittancecorrection2} are strictly valid for $\bar{\nu}_s\ll 1 $, they look very accurate also for $\bar{\nu}_s\lesssim 2.5  $, where a fraction of $1-e^{-\bar{\nu}_s}\simeq 90\%$ of the ions lying on the pulse axis will be further ionised (see Fig. \ref{fig:wholebunch}). Inspection of the saturation corrections with larger saturation parameters could be operated either by using results from Eq. \ref{eq:I} or by using numerical integration of Eq. \ref{eq:rmsbeamsat}.

\section{Summary}
We reported on a comprehensive analysis of the 3D phase-space of the particles extracted via tunnelling ionisation by a single, linearly polarised, Gaussian laser pulse. Results concerning a single-cycle averaging, showed that the model distribution of Eq. \ref{eq:model} very accurately described the distribution of the  momenta for a single ionisation process (\eg\, $Kr^{8^+\rightarrow 9^+}$). We firstly reported an estimate of the \rms \, residual momentum for the electrons extracted in a single pulse cycle. Such an estimate, valid in the limit of unsaturated ionisation, had accuracy ${\cal O}(\rho_0^2)$, \ie\, ${\cal O}(\Delta^4)$ using notation of \cite{schroeder2014thermal}, is linked to the presence of non-Gaussian terms in the extraction phase $\xi_e$ distribution (see the last raw in Eq. \ref{Wxi}). As the pulse amplitude increases approaching the saturation limit, the analysis of such a momenta distribution reveals the appearance of non-null average momentum along the single pulse peaks and a {\it decrease} of the cycle \rms\, momentum in the saturation regime. The extension of the model up two ionisation processes (\eg\, $Kr^{8^+\rightarrow 10^+}$, see also Eq. \ref{eq:rateeq_twolevels} and subsequent equations in the subsection), together with Eq. 1 gives us the possibility to predict with unprecedented accuracy the whole ionisation process occurring in a single pulse cycle. This offers either a new perspective to  analyse and prepare experiments with few-cycle pulses or a very accurate basis to simulate the cycle-averaged phase space of the extracted particles in fast codes using the {\it envelope approximation}.          

As a second outcome, we obtained a very accurate estimate of the whole bunch emittance, \ie\, the normalised emittance along the polarisation axis of the electron bunch just after the pulse passage (see Eqs.  \ref{eq:beamemittancenosat} and \ref{eq:beamemittancecorrection} for the unsaturated case and Eqs. \ref{eq:beamemittancesisat} and \ref{eq:beamemittancecorrection2} for the saturated case). Our results for the whole bunch confirmed the emittance increase in the saturation regime as firstly reported in \cite{schroeder2014thermal}, improving the results shown there by giving analytical estimates of the \rms\, transverse size increase and \rms\, momentum {\it slight decrease} due to saturation effects. 

The accuracy of the results reported in the manuscript has been checked either via full-PIC simulations or with ad-hoc Monte Carlo codes, showing a remarkable high accuracy (with errors below $1\%$) of the analytical outcomes in the fully-saturated regimes explored in the text. Our results, however, do not include the effect of the plasma wakefield where the extracted particles would be trapped. Also, transverse ponderomotive effects have not been taken into account in the analytical results concerning the transverse momentum and position separately, though their combination through the normalised emittance is not affected by the (leading term) radially linear ponderomotive force, as confirmed by our simulations.

\section*{Acknowledgments}
We acknowledge financial contribution from the CNR funded Italian Research Network ELI-Italy (D.M. No.631 08.08.2016) and from the EU Horizon 2020 Research and Innovation Program under Grant Agreement No.653782 EuPRAXIA.
The authors also wish to thank the engineers of the LLR HPC clusters and of the cluster Ruche in the Moulon Mesocentre for computer resources and help. 

\section*{Appendix}

\subsection{Optimal working point for the ionisation process}
\begin{figure}[ht]
\centering
\includegraphics[width=\linewidth]{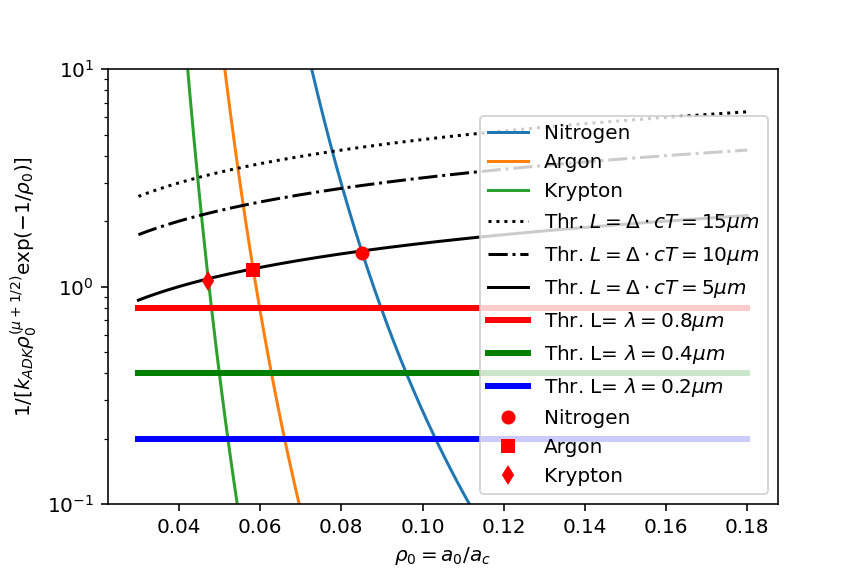}
\caption{Scale-length in $\mu m$ for ionization saturation as a function of the normalised field strength $\rho_0=a_0/a_c$  and for the Kr$^{8^+\rightarrow9^+}, (m=0)$ (green line),
Ar$^{8^+\rightarrow9^+}, (m=0)$ (orange line) and
N$^{5^+\rightarrow6^+}, (m=0)$  (light blue line) processes. The horizontal lines show the saturation point in a single cycle, while the black lines are related to long pulses of length $cT$. The red markers show the working points bringing to saturation with pulses having longitudinal size of about $5 \mu m$.  
}
\label{fig:saturation}
\end{figure}
In the special case of a single or few cycle pulse (see \cite{Lifschitz2012,Faure2019} for recent applications to the ionization injection \cite{pak2010}), a key parameter is the normalised field strength bringing to saturation into a {\it single pulse cycle}. In this case $L\approx\lambda_0$. However, in the usual case of  a long laser pulse ($cT\gg \lambda_0$), following Eq. 5 in \cite{schroeder2014thermal}, we expect a longitudinal ionization length of $L\simeq \Delta cT=\rho_0^{1/2}cT$.  We can visualize solutions of Eq. \ref{eq:satpar} by means of Fig. \ref{fig:saturation}, where the ionization scale-lengths $K(\rho_0)=[\bar{k}_{ADK} \rho_0^{\mu+1/2} e^{-1/\rho_0}]^{-1}$ and ionization lengths $L(\rho_0)$  are shown as a function of the normalised field strength $\rho_0$.

For each process, the working point is found as the intersection of the $K$ and $L$ curves in Fig. \ref{fig:saturation}.  There we show the working points realising saturation in a single wavelength for the cases $\lambda_0=0.2-0.8$ $\mu m$ (horizontal lines) or saturation for a long pulse having length in the range $5-15$ $\mu m$ (black lines). Finally, the red marks show the selected working points for pulses of length of about $cT=5$ $\mu m$.   Inspection of Fig. \ref{fig:saturation} shows that the interval of normalised field strengths of interest is very tiny. For krypton, the value of $\rho_0=0.052$ is enough to fully ionise the available ions within a single cycle with $\lambda_0=0.2$ $\mu m$. For a very long pulse with $cT=15$ $\mu m$, however, ions are close to saturation with $\rho_0=0.045$. Similarly, the field amplitudes range for argon and nitrogen are $0.055-0.065$ and $0.078-0.102$, respectively.   
\subsection*{Set-up for the PIC simulations of single-cycle ionization}
We report here the set-up of the PIC simulations with the code \Smilei\thinspace  \cite{Smilei2018,Beck2019} used to obtain Figs. \ref{fig:Argonsinglecycle}, \ref{fig:Argon2channels} .
For these simulations the azimuthal decomposition technique in cylindrical geometry has been used, with 2 azimuthal modes \cite{Lifschitz2009,Zemzemi2020,ZemzemiPhDThesis}. 
The longitudinal and radial resolutions are $\Delta z= 0.003125$ $\mu$m and $\Delta r= 0.1$ $\mu$m respectively, the integration timestep $\Delta t=0.99\thinspace\Delta z$/c.
A laser pulse with Gaussian envelope and temporal profile propagating in the $z$ direction is initialized in the simulation domain using the electromagnetic fields expressions in \cite{Quesnel1998}, multiplied by the appropriate gaussian temporal envelope. The laser pulse, with carrier wavelength $\lambda_{0}=0.4$ $\mu$m and polarized in the $x$ direction, has a waist $w_0=10$ $\mu$m and FWHM duration in intensity $T_{FWHM,d}=10$ fs, with $a_0$ taking the values for the respective simulations shown in the mentioned Figures.
The cylindrical plasma target, composed of already ionized Ar8+ and the neutralizing electrons obtained through ionization of the first 8 levels, has uniform atomic density of $10^{20}$ cm$^{-3}$, length $L_{target} = 6\Delta z$ and radius $R_{target}=8\Delta r$. Each species (ions and neutralizing electrons) of the target is sampled with $n_z\cdot n_r\cdot n_\theta=256$ macro-particles per cell, distributed regularly with $[n_z=4,n_r=4,n_\theta=16]$ particles along the $z$,$r$ directions and in the $2\pi$ azimuthal angle respectively. 
The laser pulse is initialized with CEP phase $\pi/2$, i.e. with a zero-value of the transverse electric field in the center of the laser pulse. At $t=0$, the pulse peak is positioned at the center of the target, to reproduce the underlying assumptions of the derivations.
The ionization procedure implemented in the code uses the ADK ionization rate formula as reported in \cite{schroeder2014thermal}. 
The residual parameters of the electrons obtained through ionization are computed after the laser pulse has left the target.

\subsection*{Monte Carlo simulation}
Monte Carlo simulations used the rate equations Eqs.\ref{eq:ADK} to extract particles, where the local normalised field strength $\rho=\rho_0f$ included pulse envelope effects through a Gaussian profile $f(r,z-ct)=\exp(-r^2/w_0^2-(z-ct)^2/L^2)$. As the particles have been extracted, the phase extraction $\xi_e$ was collected and the residual momentum $u_x=-(a_0f)\sin(\xi_e)$ determined along with the extraction transverse position $x$. The evaluation of the residual momentum along the polarisation axis and the particle transverse position doesn't take into account the transverse ponderomotive force and we referred in the text those simulations as "without ponderomotive force effects". A Monte-Carlo including the full electron dynamics after particle extraction, \ie\, including ponderomotive force effects has also been used. In both the cases, very large temporal resolution has been employed ($c \Delta t =\lambda_0/150$) so as to accurately describe both the ionisation process and, in the second case, the subsequent particle quivering.

\bibliographystyle{unsrtnat}

			
			
		\end{document}